%% file: tplp2017.tex
\documentclass{tlp}
\usepackage[T1]{fontenc}
\usepackage{microtype}
\usepackage{amsmath}
\usepackage{amssymb}
\usepackage{bm}
\usepackage{textcomp}
\usepackage[subrefformat=parens,labelformat=parens,format=hang,justification=raggedright]{subfig}
\usepackage{graphicx}
\usepackage{paralist}
\usepackage{listings}
\usepackage{multicol}
\usepackage[braket]{qcircuit}
\usepackage[bookmarksopen]{hyperref}
\usepackage[nameinlink]{cleveref}
\usepackage{hyperxmp}

\title
  [Fully Parallel Constraint Logic Programming on a Quantum Annealer]
  {Performing Fully Parallel Constraint Logic Programming on a Quantum Annealer}
\author[S. Pakin]{%
  SCOTT PAKIN \\
  Computer, Computational, and Statistical Sciences Division \\
  Los Alamos National Laboratory \\
  MS B287 \\
  Los Alamos, New Mexico, USA \\
  E-mail: pakin@lanl.gov
}
\hypersetup{%
  pdftitle={Performing Fully Parallel Constraint Logic Programming on a Quantum Annealer},
  pdfauthor={Scott Pakin},
  pdfauthortitle={Scientist},
  pdfcaptionwriter={Scott Pakin},
  pdfcontactemail={pakin@lanl.gov},
  pdfcontacturl={http://ccsweb.lanl.gov/\xmptilde pakin/},
  pdfcontactaddress={MS B287; Los Alamos National Laboratory},
  pdfcontactcity={Los Alamos},
  pdfcontactregion={New Mexico},
  pdfcontactpostcode={87545},
  pdfcontactcountry={USA},
  pdfsubject={Compiling Prolog to a D-Wave quantum annealer},
  pdfkeywords={%
    quantum annealing, quantum computing, constraint logic programming,
    QA Prolog, D-Wave
  },
  pdflang={en-US}
}

\pagerange{\pageref{firstpage}--\pageref{lastpage}}
\volume{\textbf{?} (?):}
\jdate{January 2017}
\pubyear{2017}
\submitted{31 March 2017}
\revised{11 January 2018}
\accepted{???}

\crefname{table}{Table}{Tables}
\crefname{figure}{Figure}{Figures}
\crefname{section}{Section}{Sections}
\crefname{listing}{Listing}{Listings}
\crefname{equation}{Equation}{Equations}
\creflabelformat{equation}{#2#1#3}
\crefname{appendix}{}{}

\let\defn=\textit                           
\newcommand*{\dw}{
  \texorpdfstring{\mbox{D-\kern-1pt Wave}}{D-Wave}%
}
\newcommand{\gmqcs}{gate-model quantum computers}  
\newcommand*{\dwiix}{\mbox{\dw~2X}}         
\newcommand*{\dwiiq}{\mbox{\dw~2000Q}}      
\newcommand*{\qap}{QA~Prolog}               
\DeclareMathOperator*{\argmin}{arg\,min}    
\let\bool=\textsc                           
\newcommand*{\hi}{\ensuremath{h_i}}         
\newcommand*{\Jij}{\ensuremath{J_{i,j}}}    
\newcommand*{\si}{\ensuremath{\sigma_i}}    
\newcommand{\liquid}{\mbox{LIQ$Ui|\rangle$}}  

\begin{document}
\maketitle
\label{firstpage}

\input{abstract}
\input{introduction}
\input{background}
\input{related-work}
\input{implementation}
\input{evaluation}
\input{conclusions}

\bibliographystyle{acmtrans}
\bibliography{tplp2017}

\appendix
\input{example}

\end{document}

%% file: abstract.tex
\begin{abstract}
  A \emph{quantum annealer} exploits quantum effects to solve a
  particular type of optimization problem.  The advantage of this
  specialized hardware is that it effectively considers all possible
  solutions in parallel, thereby potentially outperforming classical
  computing systems.  However, despite quantum annealers having
  recently become commercially available, there are relatively few
  high-level programming models that target these devices.

  In this article, we show how to compile a subset of Prolog enhanced
  with support for constraint logic programming into a 2-local
  Ising-model Hamiltonian suitable for execution on a quantum
  annealer.  In particular, we describe the series of transformations
  one can apply to convert constraint logic programs expressed in
  Prolog into an executable form that bears virtually no resemblance
  to a classical machine model yet that evaluates the specified
  constraints in a fully parallel manner.  We evaluate our efforts on
  a 1095-qubit \dwiix\ quantum annealer and describe the approach's
  associated capabilities and shortcomings.
\end{abstract}

\begin{keywords}
  quantum annealing, quantum computing, constraint logic programming,
  Prolog, \dw
\end{keywords}

%% file: introduction.tex
\section{Introduction}
\label{sec:introduction}

\defn{Quantum
  annealing}~\cite{Kadowaki1998:quant-anneal,Farhi1998:adiabatic,Finnila1994:quant-anneal}
is a presumably weaker~\cite{Bravyi2017:stoqma}\footnote{By
  ``presumably weaker'', we mean that quantum annealing is associated
  with the StoqMA complexity class~\cite{Bravyi2006:stoqma};
  gate-model quantum computing is associated with the QMA complexity
  class; and StoqMA~$\subseteq$ QMA---with a supposition that
  StoqMA~$\subset$ QMA.} but more easily
scalable~\cite{Kaminsky2004:adiabatic} form of quantum computing than
the more traditional gate model~\cite{Feynman1986:quantum-gates}.  To
quantify what we mean by ``scalable'', the recently introduced
\dwiiq\ quantum annealer provides two thousand qubits while
state-of-the-art \gmqcs\ are just starting to reach mid-double-digit
qubit counts~\cite{Knight2017:ibm-50q}.

Programming a quantum annealer is nearly identical to solving an
optimization problem using (classical) simulated
annealing~\cite{Kirkpatrick1983:sim-anneal}.  That is, one constructs
an energy landscape via a multivariate function such that the
coordinates of the landscape's ground state (i.e.,~its lowest value)
correspond to the solution being sought.  Quantum-annealing hardware
then automatically relaxes to a solution---or one of multiple equally
valid solutions---with some probability.  The ``quantum'' in ``quantum
annealer'' refers to the use of quantum effects, most notably
\defn{quantum tunneling}: the ability to ``cut through'' tall energy
barriers to reach ground states with a higher probability than could
be expected from a classical
solution~\cite{Kadowaki1998:quant-anneal}.

The advantage of quantum annealing over classical code execution is
its abundant inherent parallelism.  A quantum annealer effectively
examines all possible inputs in parallel to find solutions to a
problem.  For NP-complete problems~\cite{Cormen2001:algorithms} this
implies a potential exponential speedup over a brute-force approach.
The catch is the ``effectively''.  Quantum annealers are fundamentally
stochastic devices.  They provide no guarantee of finding an optimal
(lowest-energy) solution.  Consequently, they can be considered more
as an automatic heuristic-finding mechanism than as a formal solver.

The question we ask in this work is, \textit{Can one express
  constraint logic programming in the form accepted by
  quantum-annealing hardware?}  Although such hardware is in its first
few generations and not yet able to compete performance-wise with
traditional, massively parallel systems, our belief is that the
potential exists to one day be able to solve constraint logic
programming (CLP) problems faster on quantum annealers than on
conventional hardware.  The primary challenge lies in how to express
CLP---or, for that matter, almost any programming model---as an energy
landscape whose ground state corresponds to a satisfication of the
given constraints.  The primary contribution of this work is therefore
the demonstration that such problem expressions are indeed possible
and the presentation of a methodology (reified in software) to
accomplish that task.

The rest of this article is structured as follows.
\Cref{sec:background} details how problems need to be formulated for
execution on a quantum annealer.  Although mapping constraint logic
programs onto a quantum annealer is a novel endeavor,
\cref{sec:related-work} discusses other programming models that target
quantum annealers and \gmqcs.  \Cref{sec:implementation} is the core
part of the article.  It describes our implementation of
quantum-annealing Prolog, a CLP-enhanced Prolog subset and associated
compiler for exploiting the massive effective parallelism of a quantum
annealer when solving CLP problems.  Some examples and experiments are
presented in \cref{sec:evaluation}.  Finally, we draw some conclusions
from our work in \cref{sec:conclusions}.

%% file: background.tex
\section{Background}
\label{sec:background}

\subsection{Quantum annealing}
\label{sec:quantum-annealing}

A quantum annealer is a special-purpose device that finds a vector,
$\bm{\sigma}$, of \emph{spins} (Booleans, represented as $\pm 1$) that
minimize the energy of an Ising-model
Hamiltonian~\cite{Johnson2011:quant-ann}.  Quantum annealers from
\dw\ Systems, Inc.\ further restrict the Hamiltonian to being 2-local,
meaning that it can contain quadratic terms but not cubic or beyond.
The specific problem that a \dw\ system solves can be expressed as
\begin{equation}
  \argmin_{\bm{\sigma}} \; \mathcal{H}(\bm{\sigma})
  \text{, where~}
  \mathcal{H}(\bm{\sigma}) =
  \sum_{i=1}^N \hi \si +
  \sum_{i=1}^{N-1} \sum_{j=i+1}^N \Jij \sigma_i \sigma_j
  \label{eqn:hamiltonian}
\end{equation}
In the above, $\si \in \{-1, +1\}$, $\hi \in \mathbb{R}$, and $\Jij
\in \mathbb{R}$\@.  In other words, $\mathcal{H}$ is a pseudo-Boolean
function of degree~2.  Physically, the \hi\ represent the strength of
the external field applied to \si, and the \Jij\ represent the
strength of the interaction between \si\ and $\sigma_j$.  Given a set
of \hi\ and \Jij, finding the \si\ that minimize
$\mathcal{H}(\bm{\sigma})$ in \cref{eqn:hamiltonian} is an NP-hard
problem~\cite{Baharona1982:ising-np}.  Consequently, an efficient
(i.e.,~polynomial-time) classical algorithm for finding these \si\ in
the general case is expected not to exist.  The best known classical
algorithms run in exponential time, which is intractable for large
$N$.  (On a \dwiiq\ system, $N \approx 2000$.)  Nevertheless,
contemporary \dw\ systems can propose a solution in
\emph{microseconds}, which is an impressive capability.

A program for a quantum annealer is merely a list of \hi\ and
\Jij\ for \cref{eqn:hamiltonian}.  Clearly, there is a huge semantic
gap between such a list and constraint logic programming.  Perhaps
surprisingly, we show in \cref{sec:implementation} that it is indeed
possible to map CLP problems into Ising-model Hamiltonians.

An important point regarding \cref{eqn:hamiltonian} is that it
represents a \emph{classical} Hamiltonian.  In contrast to \gmqcs, in
which the programmer directly controls the application of
quantum-mechanical effects, these effects are almost entirely hidden
from the user of a quantum annealer.  Hence, the approach this paper
presents is equally applicable to classical annealers such as
Hitachi's CMOS annealer~\cite{Yamaoka2016:cmos-anneal}, Fujitsu's
Digital Annealer~\cite{Fujitsu2017:dig-anneal}, or even all-software
implementations of simulated
annealing~\cite{Kirkpatrick1983:sim-anneal}.  We focus our discussion
on quantum annealers, however, because such devices offer the
potential of converging to an optimal solution with higher probability
than can classical annealing methods~\cite{Kadowaki1998:quant-anneal}.

\subsection{\dw\ hardware}
\label{sec:d-wave}

\dw\ Systems, Inc.\ is a producer of commercial quantum annealers.
Although their hardware performs the basic quantum-annealing task
described in \cref{sec:quantum-annealing}, engineering reality imposes
a number of constraints on the specific Hamiltonians that can be
expressed:

\begin{itemize}
  \item As stated above, only 2-local Hamiltonians are supported.
    3-local Hamiltonians and beyond can be converted to 2-local
    Hamiltonians at the cost of additional spins.
  
  \item Even though, nominally, $\hi, \Jij \in \mathbb{R}$, those
    coefficients in fact have finite precision and are limited to
    relatively few distinct values.

  \item Although, ideally, there should be a \Jij\ for any $1 \leq
    i < j \leq N$, in practice, the system's physical topology, called
    a \defn{Chimera graph}~\cite{Bunyk2014:dwave-arch}, provides
    limited connectivity: at most six \Jij\ for any given $i$.

  \item On top of the preceding constraint, in any given installation,
    a fraction of the \hi\ and \Jij\ will be inoperative.  (See
    below.)
\end{itemize}

\Cref{fig:ising-topo} illustrates the physical topology of Ising, the
\dwiix\ system installed at Los Alamos National Laboratory (LANL) that
was used for all of the experiments reported in this article.
\begin{figure}
  \centering
  \includegraphics[width=0.75\linewidth]{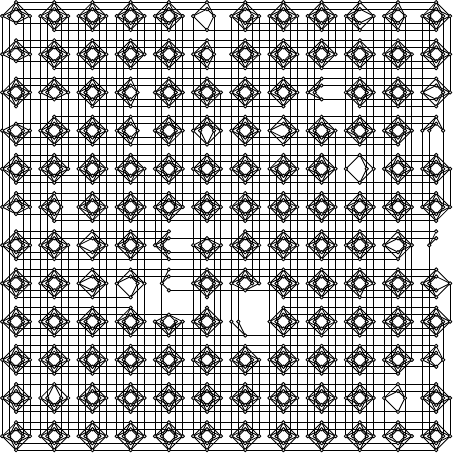}
  \caption{Physical topology of LANL's \dwiix}
  \label{fig:ising-topo}
\end{figure}
\unskip
Edges represent the \Jij, and nodes represent the \hi\ (and \si).
Physically, graph nodes are superconducting flux qubits, implemented
as niobium rings that are written and read
electromagnetically~\cite{Johnson2011:quant-ann}.  At superconducting
temperatures---Ising operates at a mere 10.45\,mK
(0.01\textcelsius\ above absolute zero)---quantum-mechanical effects
(entanglement, superposition, and quantum tunneling) come into play.
In Ising, 1095 (95.1\%) of the qubits and 3061 (91.1\%) of the
couplers are operational.

Annealing times are also installation-dependent.  Ising supports
annealing times of 5--2000\textmu{}s (user-selectable).  Longer
annealing times are theoretically more likely to find the input
Hamiltonian's global minimum, but shorter annealing times enable more
attempts per unit time.

%% file: related-work.tex
\section{Related Work}
\label{sec:related-work}

Very little exists in terms of programming models and programming
languages for \gmqcs\ and quantum annealers.

\begin{figure}
  \centering
  \mbox{}%
  \Qcircuit @C=1.25em @R=1.25em {
    \lstick{\ket{0}} & \qw      & \gate{H} & \gate{X} & \ctrl{2} & \gate{H} & \gate{X} & \qw      & \ctrl{1} & \qw      & \gate{X} & \gate{H} & \meter & \qw \\
    \lstick{\ket{0}} & \qw      & \gate{H} & \qw      & \ctrl{1} & \gate{H} & \gate{X} & \gate{H} & \targ    & \gate{H} & \gate{X} & \gate{H} & \meter & \qw \\
    \lstick{\ket{0}} & \gate{X} & \gate{H} & \qw      & \targ    & \qw      & \qw      & \qw      & \qw      & \qw      & \qw      & \qw      & \qw    & \qw
    \gategroup{1}{4}{3}{5}{0.5em}{--}
  }
  \caption{Sample gate-model program (Grover's method)}
  \label{fig:grover}
\end{figure}
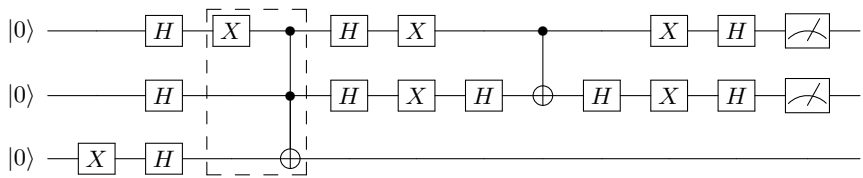

\paragraph{Programming models for \gmqcs}
Although \gmqcs\ are more heavily studied than quantum annealers,
virtually all programming languages and compilers developed for these
systems are based on the same underlying programming model: Place a
specified gate at a specified location in a circuit, which is a
directed acyclic graph.  To illustrate this approach,
\cref{fig:grover} presents an example of Grover's search
algorithm~\cite{Grover1996:search} in standard gate-model circuit
notation.\footnote{The dashed box delineates the user-provided
  ``oracle'' function, which indicates if a set of inputs (2~bits, in
  this case) represents the item being searched for by flipping an
  ancilla qubit.}  Each gate (e.g.,~$X$, $H$, or \textit{CNOT}
[$\bullet\text{---}\mathord{\bigoplus}$]) represents a small unitary
matrix that transforms the current state---in this case, a complex
vector with 8~($2^3$) elements.

The question is how to express algorithms such as the one shown in
\cref{fig:grover} as computer code.
OpenQASM~\cite{Cross2017:openqasm} is a domain-specific language in
which the ``statements'' (\texttt{x}, \texttt{h}, \texttt{cx}, etc.)
represent the application of a gate to one or more qubits, and
programs can define parameterized macros to simplify repeated tasks.
Scaffold~\cite{JavadiAbhari2015:scaffcc} is similar but also supports
C-style \texttt{for} loops to apply multiple gates in a fixed pattern.
It further includes support for expressing classical oracles as gates.
Quipper~\cite{Green2013:quipper} and \liquid~\cite{Wecker2014:liquid}
are embedded domain-specific languages, with the former embedded in
Haskell and the latter embedded in F\#, implying they have access to
those languages' code features.  Like OpenQASM and Scaffold, Quipper
and \liquid\ are based primarily on specifying an ordered sequence of
gates applied to qubits, although in a functional-language context
(e.g.,~a Quipper circuit runs within a Haskell monad).
Quil~\cite{Smith2017:quil} and Q\#~\cite{Microsoft2017:q-sharp} are
recent domain-specific languages that, while sharing the same
``ordered sequence of gates'' abstraction as all the other efforts,
put particular emphasis on tightly interleaving classical and quantum
computation.

In short, the state of the art in programming models for \gmqcs\ is at
the level of an assembly language that offers various convenience
features but little in the way of higher-level abstractions.  The
closest equivalent in the context of quantum annealing is perhaps
QMASM~\cite{Pakin2016:qmasm}, which we introduce in
\cref{sec:qa-prolog} but then build upon to present a more expressive
programming model based on classical circuits and, on top of that, one
based on predicate logic.  Unlike all of the works discussed above, in
this paper we establish a large semantic gap between the programming
model exposed by the underlying hardware and that presented to the
user.

\paragraph{Quantum Prolog}
\citeNS{James2011:qc-prolog} demonstrate that one can express the
equivalent of a pure version of Prolog over finite relations in terms
of a model of \emph{discrete} quantum computing.  (Conventionally,
quantum computing is defined over Hilbert spaces, which are
complex-valued.)  This is one of the very few attempts to develop a
high-level programming model for \gmqcs, although for a form that has
never been implemented.  Besides targeting \gmqcs\ rather than quantum
annealers, this work differs from ours in that it focuses on the
\emph{mathematical equivalency} of relational programming and discrete
quantum computing over the field of Booleans, while our work showcases
the \emph{implementation} of a Prolog compiler that generates code
suitable for running on a physical quantum annealer.

\paragraph{Programming models for quantum annealers}
Most of the literature that relates to programming quantum annealers
focuses on expressing a single algorithm or single class of algorithms
in terms of an Ising-model Hamiltonian (\cref{eqn:hamiltonian}).
\citeN{Lucas2014:np-probs} surveys a number of such algorithms.  More
recent efforts include traveling-salesman
problems~\cite{Heim2017:adiabatic-tsp} and satisfiability
problems~\cite{Hen2011:exp-complex}.  In contrast, our work is to make
it possible to express problems without explicitly specifying
individual \hi\ and \Jij\ coefficients.

\dw\ Systems, Inc.\ provides a few tools that let one express problems
in a higher-level form than a list of \hi\ and \Jij\ coefficients.
ToQ~\cite{D-Wave:toq}, formerly Deqo~\cite{Dahl2014:deqo}, is the most
related tool to our quantum-annealing Prolog in that it is centered on
constraint satisfaction and targets a quantum annealer.  ToQ accepts a
set of constraints and returns a set of values that satisfy those
constraints.  However, each individual constraint is evaluated
classically and exhaustively; the \dw\ system is used only for
combining individually satisfied constraints into a global problem and
solving that.  In contrast, our Prolog implementation performs
\emph{all} of its constraint satisfaction on the quantum annealer, not
classically.

%% file: implementation.tex
\section{Implementation}
\label{sec:implementation}

\subsection{Primitives}
\label{sec:primitives}

\Cref{tbl:h-effect,tbl:J-effect} present the primitive mechanisms that
one can use to represent a problem as an Ising-model Hamiltonian of
the form stated in \cref{eqn:hamiltonian}.
\begin{table}
  \caption{Effect of negative and positive \hi}
  \label{tbl:h-effect}
  \centering
  \subfloat[$\hi < 0$: Favor \bool{true}\label{tbl:h-effect-1}]{%
    \begin{minipage}{9pc}
      \begin{tabular}{@{}rrc@{}}
        \hline\hline
        \si & $-1 \, \si$ & $\argmin_{\bm{\sigma}}$ \\
        \hline
        $-1$ & $+1$ & \\
        $+1$ & $-1$ & \checkmark \\
        \hline\hline
      \end{tabular}
    \end{minipage}}
  \hfil
  \subfloat[$\hi > 0$: Favor \bool{false}\label{tbl:h-effect+1}]{%
    \begin{minipage}{9pc}
      \begin{tabular}{@{}rrc@{}}
        \hline\hline
        \si & $+1 \, \si$ & $\argmin_{\bm{\sigma}}$ \\
        \hline
        $-1$ & $-1$ & \checkmark \\
        $+1$ & $+1$ & \\
        \hline\hline
      \end{tabular}
    \end{minipage}}
\end{table}
\begin{table}
  \caption{Effect of negative and positive \Jij}
  \label{tbl:J-effect}
  \centering
  \subfloat[$\Jij < 0$: Favor equality\label{tbl:J-effect-1}]{%
    \begin{minipage}{12pc}
      \begin{tabular}{@{}rrrc@{}}
        \hline\hline
        \si & $\sigma_j$ & $-1 \, \sigma_i \sigma_j$ & $\argmin_{\bm{\sigma}}$ \\
        \hline
        $-1$ & $-1$ & $-1$ & \checkmark \\
        $-1$ & $+1$ & $+1$ &            \\
        $+1$ & $-1$ & $+1$ &            \\
        $+1$ & $+1$ & $-1$ & \checkmark \\
        \hline\hline
      \end{tabular}
    \end{minipage}}
  \hfil
  \subfloat[$\Jij > 0$: Favor inequality\label{tbl:J-effect+1}]{%
    \begin{minipage}{12pc}
      \begin{tabular}{@{}rrrc@{}}
        \hline\hline
        \si & $\sigma_j$ & $+1 \, \sigma_i \sigma_j$ & $\argmin_{\bm{\sigma}}$ \\
        \hline
        $-1$ & $-1$ & $+1$ &            \\
        $-1$ & $+1$ & $-1$ & \checkmark \\
        $+1$ & $-1$ & $-1$ & \checkmark \\
        $+1$ & $+1$ & $+1$ &            \\
        \hline\hline
      \end{tabular}
    \end{minipage}}
\end{table}
\unskip
Interpreting $\si = -1$ as \bool{false} and $\si = +1$ as \bool{true},
a negative \hi\ expresses a preference for the corresponding
\si\ being \bool{true} (\cref{tbl:h-effect-1}) while a positive
\hi\ expresses a preference for the corresponding \si\ being
\bool{false} (\cref{tbl:h-effect+1}).  The magnitude of the
\hi\ corresponds to the strength of the preference.  A negative
\Jij\ expresses a preference for the corresponding \si\ and $\sigma_j$
having the same value (\cref{tbl:J-effect-1}) while a positive
\Jij\ expresses a preference for the corresponding \si\ and $\sigma_j$
having opposite values (\cref{tbl:J-effect+1}).  In other words, $\Jij
< 0$ can be interpreted as a wire in a digital circuit, and $\Jij > 0$
can be interpreted as an inverter.  As with the \hi, the magnitude of
the \Jij\ corresponds to the strength of the preference.

One can set up a system of inequalities to find \hi\ and \Jij\ values
with a desired set (or sets) of \si\ in the ground state.  This system
of inequalities can be solved by hand in simple cases or with a
constraint solver for more complicated cases.  We use
MiniZinc~\cite{Nethercote2007:minizinc} as our constraint-modeling
language, but any similar system would suffice.
\Cref{tbl:simple-bools-and} presents an Ising-model Hamiltonian,
$\mathcal{H}_\wedge(\bm{\sigma})$, whose 4-way degenerate ground state
(meaning, the 4-way tie for the minimum value) is exactly the set of
spins for which $\sigma_k = \sigma_i \wedge \sigma_j$ (i.e.,~logical
conjunction).
\begin{table}
  \caption{Representing simple Boolean functions as Ising-model Hamiltonians}
  \label{tbl:simple-bools}
  \subfloat[Logical conjunction (\bool{and})\label{tbl:simple-bools-and}]{%
    \begin{minipage}{13pc}
      \begin{tabular}{@{}rrrrc@{}}
        \hline\hline
        $\sigma_i$ & $\sigma_j$ & $\sigma_k$ & $\mathcal{H}_\wedge(\bm{\sigma})$ & $\argmin_{\bm{\sigma}}$ \\
        \hline
        $-1$ & $-1$ & $-1$ & $-1.5$ & \checkmark \\
        $-1$ & $-1$ & $+1$ & $ 4.5$ &            \\
        $-1$ & $+1$ & $-1$ & $-1.5$ & \checkmark \\
        $-1$ & $+1$ & $+1$ & $ 0.5$ &            \\
        $+1$ & $-1$ & $-1$ & $-1.5$ & \checkmark \\
        $+1$ & $-1$ & $+1$ & $ 0.5$ &            \\
        $+1$ & $+1$ & $-1$ & $ 0.5$ &            \\
        $+1$ & $+1$ & $+1$ & $-1.5$ & \checkmark \\
        \hline\hline
      \end{tabular}
      $\mathcal{H}_\wedge(\bm{\sigma}) = -0.5\sigma_i + -0.5\sigma_j + 1.0\sigma_k
      + 0.5\sigma_i\sigma_j + -1.0\sigma_i\sigma_k + -1.0\sigma_j\sigma_k$
    \end{minipage}}
  \hfil
  \subfloat[Logical disjunction (\bool{or})\label{tbl:simple-bools-or}]{%
    \begin{minipage}{13pc}
      \begin{tabular}{@{}rrrrc@{}}
        \hline\hline
        $\sigma_i$ & $\sigma_j$ & $\sigma_k$ & $\mathcal{H}_\vee(\bm{\sigma})$ & $\argmin_{\bm{\sigma}}$ \\
        \hline
        $-1$ & $-1$ & $-1$ & $-1.5$ & \checkmark \\
        $-1$ & $-1$ & $+1$ & $ 0.5$ &            \\
        $-1$ & $+1$ & $-1$ & $ 0.5$ &            \\
        $-1$ & $+1$ & $+1$ & $-1.5$ & \checkmark \\
        $+1$ & $-1$ & $-1$ & $ 0.5$ &            \\
        $+1$ & $-1$ & $+1$ & $-1.5$ & \checkmark \\
        $+1$ & $+1$ & $-1$ & $ 4.5$ &            \\
        $+1$ & $+1$ & $+1$ & $-1.5$ & \checkmark \\
        \hline\hline
      \end{tabular}
      $\mathcal{H}_\vee(\bm{\sigma}) = 0.5\sigma_i + 0.5\sigma_j + -1.0\sigma_k
      + 0.5\sigma_i\sigma_j + -1.0\sigma_i\sigma_k + -1.0\sigma_j\sigma_k$
    \end{minipage}}
\end{table}
The Hamiltonian was found by constraining the ground-state
$\mathcal{H}_\wedge(\bm{\sigma})$ all to have the same value and all
other $\mathcal{H}_\wedge(\bm{\sigma})$ to have a strictly greater
value.  Similarly, \cref{tbl:simple-bools-or} presents an Ising-model
Hamiltonian, $\mathcal{H}_\vee(\bm{\sigma})$, whose 4-way degenerate
ground state is exactly the set of spins for which $\sigma_k =
\sigma_i \vee \sigma_j$ (i.e.,~logical disjunction).

It is worth noting that Hamiltonians are additive.  Specifically, the
ground state of $\mathcal{H}_A + \mathcal{H}_B$ is the intersection of
the ground state of $\mathcal{H}_A$ and the ground state of
$\mathcal{H}_B$.  The implication is that primitive operations like
those shown in \cref{tbl:simple-bools} can be trivially combined into
more complicated expressions without having to solve a
(computationally expensive) constraint problem for the overall
expression.  Because Ising-model Hamiltonians always have a non-empty
ground state---they must have at least one minimum value---summing
Hamiltonians whose ground states have an empty intersection can lead
to an unintuitive ground state of the combined Hamiltonian.  As a
trivial example, $\mathcal{H}_A = \sigma_a - \sigma_b$ has the unique
ground state $\{\sigma_a = -1, \sigma_b = +1\}$; $\mathcal{H}_B =
-\sigma_a \sigma_b$ has the two-fold degenerate ground state
$\{\sigma_a = -1, \sigma_b = -1\}$ and $\{\sigma_a = +1, \sigma_b =
+1\}$; but $\mathcal{H}_A + \mathcal{H}_B$ has the \emph{three-fold}
degenerate ground state $\{\sigma_a = -1, \sigma_b = -1\}$,
$\{\sigma_a = -1, \sigma_b = +1\}$, and $\{\sigma_a = +1, \sigma_b =
+1\}$.

In this work, however, we ensure by construction that we do not sum
Hamiltonians whose intersection is empty.  Specifically, because we
sum Hamiltonians only for Boolean expressions and only by equating
(cf.~\cref{tbl:J-effect-1}) outputs to inputs and never outputs to
outputs or inputs to inputs (i.e.,~we rely on a directed acyclic graph
organization) we will not normally wind up with an empty intersection.
The exception is in the case in which spins are ``pinned'' to values
that introduce unsatisfiable output-output or input-input couplings,
as discussed on \cpageref{page:pinning} in the following section.

Armed with Hamiltonians for \bool{and} (\cref{tbl:simple-bools-and}),
\bool{or} (\cref{tbl:simple-bools-or}), and \bool{not}
(\cref{tbl:J-effect+1}) and the knowledge that Hamiltonians can be
added together to produce new, more constrained Hamiltonians, we can
implement a complete zeroth-order logic.

\subsection{QA Prolog}
\label{sec:qa-prolog}

We have implemented a Prolog compiler that compiles Prolog programs to
a list of \hi\ and \Jij\ coefficients for use with
\cref{eqn:hamiltonian}.  We call our implementation
\defn{quantum-annealing Prolog} or \defn{QA Prolog} for short.
Although the underlying concept is no more sophisticated than what was
described in \cref{sec:primitives}, a large software-engineering
effort was needed to bridge the gap between what was presented there
and a usable Prolog compiler.

Internally, QA Prolog comprises a number of software layers.  The
lowest layer is a ``quantum macro assembler'' we developed called
QMASM~\cite{Pakin2016:qmasm}.\footnote{QMASM is freely available from
  \url{https://github.com/lanl/qmasm}.}  QMASM provides a thin but
convenient layer of abstraction atop \cref{eqn:hamiltonian}.  It lets
programs refer to spins symbolically rather than numerically; shields
the program from having to consider the specific underlying physical
topology; and provides modularization through the use of macros that
can be defined once and instantiated repeatedly.

\newsavebox{\qmasmOR}
\begin{lrbox}{\qmasmOR}
  \begin{minipage}[t]{0.45\linewidth}
    \figrule
    \begin{lstlisting}[
      alsoletter=!,
      morekeywords={!begin_macro,!end_macro,!use_macro,!include},
      morecomment={[l]\#},
      breaklines
    ]
!begin_macro or
A  0.5
B  0.5
Y -1.0

A B  0.5
A Y -1.0
B Y -1.0
!end_macro or
    \end{lstlisting}
    \figrule
  \end{minipage}
\end{lrbox}

\newsavebox{\qmasmAND}
\begin{lrbox}{\qmasmAND}
  \begin{minipage}[t]{0.45\linewidth}
    \figrule
    \begin{lstlisting}[
      alsoletter=!,
      morekeywords={!begin_macro,!end_macro,!use_macro,!include},
      morecomment={[l]\#},
      breaklines
    ]
!begin_macro and
A -0.5
B -0.5
Y  1.0

A B  0.5
A Y -1.0
B Y -1.0
!end_macro and
    \end{lstlisting}
    \figrule
  \end{minipage}
\end{lrbox}

As an example, \cref{fig:qmasm-and-or} shows how one can define
\texttt{and} and \texttt{or} macros corresponding to the Hamiltonians
portrayed by \cref{tbl:simple-bools}.
\begin{figure}
  \centering
  \subfloat[QMASM version of \cref{tbl:simple-bools-and}\label{fig:qmasm-and}]{%
    \usebox{\qmasmAND}}
  \hfil
  \subfloat[QMASM version of \cref{tbl:simple-bools-or}\label{fig:qmasm-or}]{%
    \usebox{\qmasmOR}}
  \caption{QMASM macro definitions for \bool{and} and \bool{or}}
  \label{fig:qmasm-and-or}
\end{figure}
Lines containing a single symbol and a value correspond to an \hi, and
lines containing two symbols and a value correspond to a \Jij.  Macros
can be instantiated using the \textbf{!use\_macro} directive.

The QMASM code in \cref{fig:qmasm-and,fig:qmasm-or} is compiled to a
physical Hamiltonian, i.e.,~one that uses only the \hi\ and \Jij\ that
exist on the specific underlying hardware, with all coefficients
scaled to the supported range.  The \dw's physical, Chimera-graph
topology is not only fairly sparse but also contains no odd-length
cycles---needed for the two A--B--Y cycles in \cref{fig:qmasm-and-or},
for example---implying that a typical Hamiltonian must be
\defn{embedded}~\cite{Choi2008:minor-embed} into the physical
topology.  Doing so requires that additional spins and additional
terms be added to the Hamiltonian and slightly alters the
coefficients.  For example, \cref{fig:qmasm-and} may compile to
$\mathcal{H}_\wedge(\bm{\sigma}) = -0.125\sigma_8 - 0.25\sigma_9 +
0.5\sigma_{14} - 0.125\sigma_{15} - 0.5\sigma_8\sigma_{14} -
0.5\sigma_9\sigma_{14} - 1.0\sigma_8\sigma_{15} +
0.25\sigma_9\sigma_{15}$.  A benefit of QMASM is that it lets programs
work with arbitrary 2-local Ising Hamiltonians---support for
higher-order interaction terms may be added in the future---while it
automatically maps those Hamiltonians onto the available hardware.

Given that we can implement Boolean functions as Hamiltonians, we can
take a large leap in abstraction and programmability and map Verilog
programs~\cite{Thomas2002:verilog} into the form of
\cref{eqn:hamiltonian}.  Verilog is a popular hardware-description
language.  Unlike QMASM, which looks foreign to a
conventional-language programmer, the Verilog language supports
variables, arithmetic operators, conditionals, loops, and other common
programming-language constructs.

Verilog is a good match for current quantum annealers because, unlike
most programming languages, it provides precise control over the
number of bits used by each variable.  With a total of only a few
thousand bits (a few hundred bytes) available for both code and data
\emph{combined} on contemporary quantum annealers, there is no room
for waste.  \Cref{fig:verilog-example} presents an example of a
Verilog module that inputs two 4-bit variables, \textit{this} and
\textit{that}, and outputs a 5-bit variable, \textit{the\_other},
which is a function of the two inputs.
\begin{figure}
  \centering
  \begin{minipage}[t]{20pc}
    \figrule
    \begin{lstlisting}[
        language=Verilog,
        columns=fullflexible,
        breaklines
      ]
module silly (this, that, the_other);
   input [3:0] this, that;
   output [4:0] the_other;
   wire [4:0] this2, that2;

   assign this2 = (this[0] == 1'b1) ? this*5'd2 : this;
   assign that2 = (that[0] == 1'b1) ? that*5'd2 : that;
   assign the_other = this2 + that2;
endmodule
    \end{lstlisting}
    \figrule
  \end{minipage}
  \caption{Sample Verilog program}
  \label{fig:verilog-example}
\end{figure}
Although this program does not perform a useful function, it works
well for pedagogical purposes because it employs a variety of language
features including internal variables (defined with \textbf{wire}), a
relational operator (\verb|==|), a C-style ternary conditional
(\verb|?:|), multiplication, addition, and assignment.

We use Yosys~\cite{Wolf2013:yosys}, an open-source hardware-synthesis
tool, to compile Verilog code to an EDIF (Electronic Design
Interchange Format) netlist~\cite{Kahn2000:edif}, a precise,
machine-parseable description of a digital circuit.  In addition to
compiling, Yosys performs a number of optimizations on the design and,
with the help of ABC~\cite{Brayton2010:abc,Berkeley2006:abc},
transforms the design to use only a relatively small set of basic
gates.  A program we developed, edif2qmasm,\footnote{edif2qmasm is
  freely available from \url{https://github.com/lanl/edif2qmasm}.}
translates the EDIF netlist to a QMASM Hamiltonian.  The generated
code employs a standard-cell library of precomputed Hamiltonians for
\bool{and}, \bool{or}, \bool{not}, \bool{xor}, and other such
primitives.

In the case of \cref{fig:verilog-example}, the generated Hamiltonian
comprises 263~\si\ and \hi\ and 375~\Jij.  We can
``pin''\label{page:pinning} values to the \textit{this} and
\textit{that} inputs by specifying the \hi\ as in \cref{tbl:h-effect},
run the Hamiltonian on a quantum annealer, and read out the
\textit{the\_other} output.  (QMASM even supports pinning directly on
the command line~\cite{Pakin2016:qmasm}.)  We can alternatively pin
the \textit{the\_other} output, run the Hamiltonian, and read out the
two inputs; pin the output and one of the inputs and read out the
other input; or pin nothing and read out valid mappings of inputs to
outputs.  (As could be expected, though, if the function were
non-surjective, specifying an output that is not the image of any
element in the function's domain will not result in valid inputs.)  In
essence, we support a relational semantics in a language that does not
normally offer such a capability.

The final piece of \qap\ is the Prolog compiler itself.
\qap\ supports only a subset of Prolog but enough to handle basic
constraint logic programming.  For instance, \qap\ supports atoms and
positive integers but not floating-point numbers, strings, lists,
first-class compound terms, or any other data type.  It supports
arithmetic and relational operations but not \texttt{!}~(cut),
\texttt{fail}, or any impure predicates.  Clauses can reference other
clauses but not recursively.  Polymorphic clauses are not supported,
but \texttt{integer/1} and \texttt{atom/1} can be used to disambiguate
otherwise polymorphic clauses.  \qap\ does support
unification~\cite{Robinson1965:unification}, backtracking, and
predicates comprising multiple clauses.

Importantly, \qap\ allows the goals in a rule's body to be specified
in any order without impacting their ability to be proven.  In
particular, operations can be performed on variables even before they
are ground.  This is in contrast to basic Prolog---as opposed to
Prolog with the \texttt{library(clpfd)}
predicates~\cite{Triska2012:clpfd}---which is limited in its ability
to manipulate free variables.

After the usual lexing and parsing steps, the compiler performs type
inference on the abstract syntax tree.  Because so few spins are
available in current hardware, distinguishing between the two
supported data types lets the compiler represent each of them using a
different numbers of bits: $\lceil \log_2(a) \rceil$ bits for atoms,
assuming $a$~distinct atoms are named in the program, and $\lceil
\log_2(n) \rceil$ bits for integers, assuming the largest integer
appearing literally in the program is~$n$.  A \qap\ command-line
argument lets the user increase the number of bits per integer in case
larger values are needed for intermediate results.

Once all types are inferred, \qap's code generator generates Verilog
code.  Although compiling Prolog to Verilog appears, on the surface,
to be a peculiar strategy, recall that
\begin{enumerate}
  \item our quantum-annealing implementation of Verilog already has a
    relational semantics,

  \item Verilog provides arithmetic and relational operations, saving
    \qap\ from having to implement those itself,

  \item hardware-synthesis tools such as Yosys perform a number of
    logic optimizations and simplifications on \qap's behalf, and

  \item \qap\ gets unification support ``for free'' because the same
    spins are used to represent every instance of the same variable
    appearing in the Verilog code.
\end{enumerate}

In the compilation from Prolog to Verilog, each predicate (including
all clauses) is converted to a single Verilog module.  Because the
names of Verilog arguments must be unique, arguments are renamed.  In
Prolog terms, this is like replacing the fact ``\texttt{same(X, X).}''
with the rule ``\texttt{same(A, B) :- A = B.}''.  In addition, an
extra, single-bit argument called \texttt{Valid} is included in the
argument list.  This is an output value that is set to~\texttt{1} if
and only if all goals are proved.  Queries that include variables are
implemented by pinning \texttt{Valid} to~\texttt{1} and letting the
annealing process solve for all the variables so as not to violate
that condition.  Queries that do not include variables leave
\texttt{Valid} unpinned and let the annealing process find a value for
it that does not violate any other conditions.

\begin{figure}
  \centering
  \includegraphics[height=0.3\textheight]{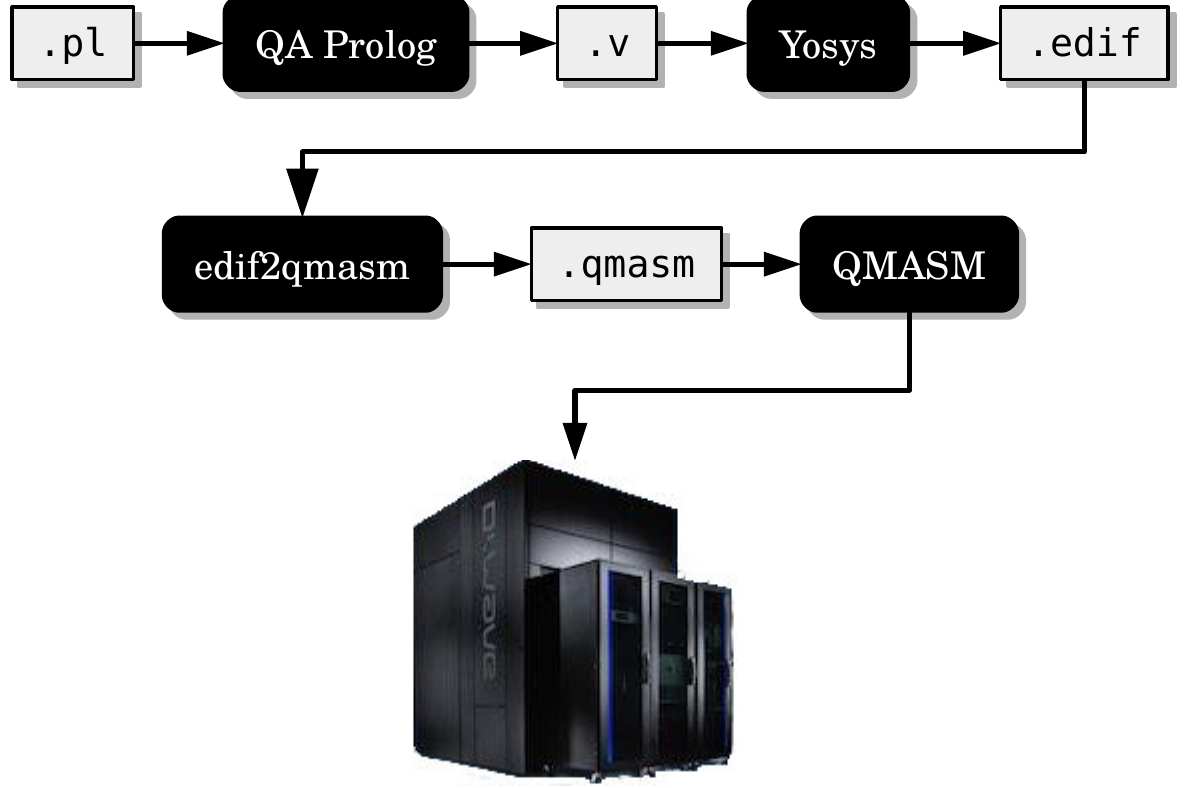}
  \caption{Overall \qap\ processing flow}
  \label{fig:qap-flow}
\end{figure}

Once \qap\ has compiled the Prolog source code---including the query,
which can be specified on the command line---to Verilog, \qap\ invokes
Yosys, edif2qmasm, and QMASM, as illustrated in \cref{fig:qap-flow}.
With the help of \dw's SAPI library~\cite{D-Wave2016:sapi-python},
QMASM remotely executes the user's program on a \dw\ system and
reports the (Boolean) value of each symbol appearing in the QMASM
source file.  \qap\ maps these lists of Booleans back to integers and
named atoms, associates those values with variables named in the
user's query, and reports all variables and their values to the user
just like a typical Prolog environment would.

A quantum annealer \emph{always} returns a vector of spins
($\bm{\sigma}$); there is no notion of ``no result found''.  Although
QMASM can detect certain obviously incorrect results and discard them,
in the general case, \qap\ can return incorrect results.  Consider the
constraint logic program, ``\texttt{impossible(X)~:- X~<~4, X~>~4.}''.
Although \texttt{impossible(X)} should never succeed, \qap\ proposes
all eight 3-bit integers as possible solutions, as all eight are
equally bad choices.  We do not consider this odd behavior a
showstopper because many important problems in~P and~NP have solutions
that can be (classically) verified quickly.  A user can run \qap\ to
quickly produce a set of candidate solutions then filter out invalid
ones as part of a post-processing step.

%% file: evaluation.tex
\section{Evaluation}
\label{sec:evaluation}

In this section, we examine what \qap\ is and is not capable of
expressing.  \Cref{sec:example} expands upon this section by
presenting all of the transformations a particular program undergoes
from Prolog source code to a 2-local Ising Hamiltonian.  Our main
metric in this section is the cost in spins for various programs.
\Cref{fig:and3} presents a Prolog program that defines an
\texttt{and/3} predicate in terms of an \texttt{and/2} predicate,
which itself is defined by four facts.
\begin{figure}
  \centering
  \begin{minipage}[t]{9pc}
    \figrule
    \begin{lstlisting}[
        columns=fullflexible,
        breaklines
      ]
and(false, false, false).
and(false, true,  false).
and(true,  false, false).
and(true,  true,  true).

and(A, B, C, Y) :-
        and(A, B, AB),
        and(AB, C, Y).
    \end{lstlisting}
    \figrule
  \end{minipage}
  \caption{3-input \bool{and}}
  \label{fig:and3}
\end{figure}
Note that this example does not involve any constraint logic
programming.

Compiling \cref{fig:and3} with \qap\ results in a logical Hamiltonian
with 26~spins, and QMASM assembles that into a physical Hamiltonian
with 42~spins.  For comparison, a hand-constructed logical Hamiltonian
for a 3-input \bool{and} requires only
5~spins,\footnote{$\mathcal{H}_{\text{and/3}}(\bm{\sigma}) = -
  \frac{1}{2} \sigma_{A} - \frac{1}{2} \sigma_{B} - \frac{1}{2}
  \sigma_{C} + \sigma_{Y} + \frac{1}{2} \sigma_{x} + \frac{1}{2}
  \sigma_{A} \sigma_{B} - \sigma_{A} \sigma_{x} - \sigma_{B}
  \sigma_{x} - \sigma_{C} \sigma_{Y} + \frac{1}{2} \sigma_{C}
  \sigma_{x} - \sigma_{Y} \sigma_{x}$, where $\sigma_x$ is an ancilla
  spin needed to produce the correct ground state.} and QMASM
assembles that into a physical Hamiltonian containing 10~spins.

Next, consider a Prolog program that finds two positive integers whose
sum and product are each~4 (\cref{fig:four}).
\begin{figure}
  \centering
  \begin{minipage}[t]{7pc}
    \figrule
    \begin{lstlisting}[
        language=Prolog,
        columns=fullflexible,
        breaklines
      ]
fours(A, B) :-
        A+B = 4,
        A*B = 4.
    \end{lstlisting}
    \figrule
  \end{minipage}
  \caption{Find two numbers that both add and multiply to~4}
  \label{fig:four}
\end{figure}
When run with the query ``\texttt{fours(A, B)}'', the program
correctly produces ``\texttt{A~= 2, B~= 2}''.  The program in fact
additionally produces ``\texttt{A~= 6, B~= 6}''.  Because 3 is the
minimum number of bits needed to represent all integers appearing
literally in \cref{fig:four}, \qap\ represents all numbers with that
many bits.  Because $6+6 \equiv 4 \pmod{2^3}$ and $6 \cdot 6 \equiv 4
\pmod{2^3}$, the second solution is valid, albeit a bit surprising.

Even a program this small consumes a large number of spins, primarily
because of the 3-bit multiplication.  Specifically, it maps to 24
logical spins at the QMASM level, which in turn get mapped to 39
physical spins or 3.6\% of the total available on LANL's
\dwiix\ system.  The number of logical spins is a function of both the
compilation tools (Yosys and ABC, in this case) and the set of
precomputed Hamiltonians included in edif2qmasm's standard-cell
library.  The number of physical spins is a function of the both the
physical topology and the minor-embedding algorithm---and the
random-number generator's state in the case of a stochastic embedder
such as the one we use~\cite{Cai2014:graph-minors}.

The total end-to-end execution time, including all of the compilation
steps and the network connection to the \dwiix, averages $3.0 \pm
1.0$\,s over 100~trials.  For each trial, we specified an annealing
time of 20\,\textmu{}s and had the hardware perform 1000~anneals,
enabling a maximum of 1000 unique solutions to be returned.  Hence, a
fixed 20\,ms of the end-to-end time is the computation proper---the
actual annealing time.  Although the end-to-end time is high, our
belief is that this time will grow slower with program complexity than
would a classical implementation.  Remember: all goals in the entire
program are effectively evaluated in parallel.  Not only are CLP
semantics honored, but there is no additional performance cost for
going beyond plain Prolog's unification capabilities.

What are the limits on the Prolog programs that can be run on a
\dwiix\ using \qap?  We found that the ``light meal'' example from
Dutra's CLP tutorial~\cite{Dutra2010:clp-tut} is too large to fit, but
if we skip dessert, as in \cref{fig:meal}, the code runs, with 170
logical spins dilating to 602 physical spins.
\begin{figure}
  \centering
  \begin{minipage}[t]{15pc}
    \figrule
    \begin{lstlisting}[
        language=Prolog,
        columns=fullflexible,
        breaklines
      ]
light_meal(S, M) :-
        Skc + Mkc =< 8,
        starter(S, Skc),
        main_course(M, Mkc).

meat(steak, 5).
meat(pork, 7).
fish(sole, 2).
fish(tuna, 4).

main_course(M, Mkc) :- meat(M, Mkc).
main_course(M, Mkc) :- fish(M, Mkc).

starter(salad, 1).
starter(soup, 6).
    \end{lstlisting}
    \figrule
  \end{minipage}
  \caption{Planning a meal of no more than 8~kcal}
  \label{fig:meal}
\end{figure}
As a more controlled experiment, consider a \texttt{mult/3} predicate
defined as ``\texttt{mult(A, B, C)~:- C~= A*B.}''.  Because of \qap's
support for CLP, we can provide a value only for \texttt{C} in a query
to factor \texttt{C} into \texttt{A} and \texttt{B}.  The reader
should note that this one-line integer-factoring program for a quantum
annealer is conceptually far simpler than Shor's famous
integer-factoring algorithm for \gmqcs~\cite{Shor1999:factoring},
which requires knowledge of both quantum mechanics and number theory
to understand.  By factoring a relatively small number, say~6, we can
steadily increase the bit width and measure the number of logical and
physical spins required to represent the program.

\begin{figure}
  \centering
  \includegraphics[scale=0.9]{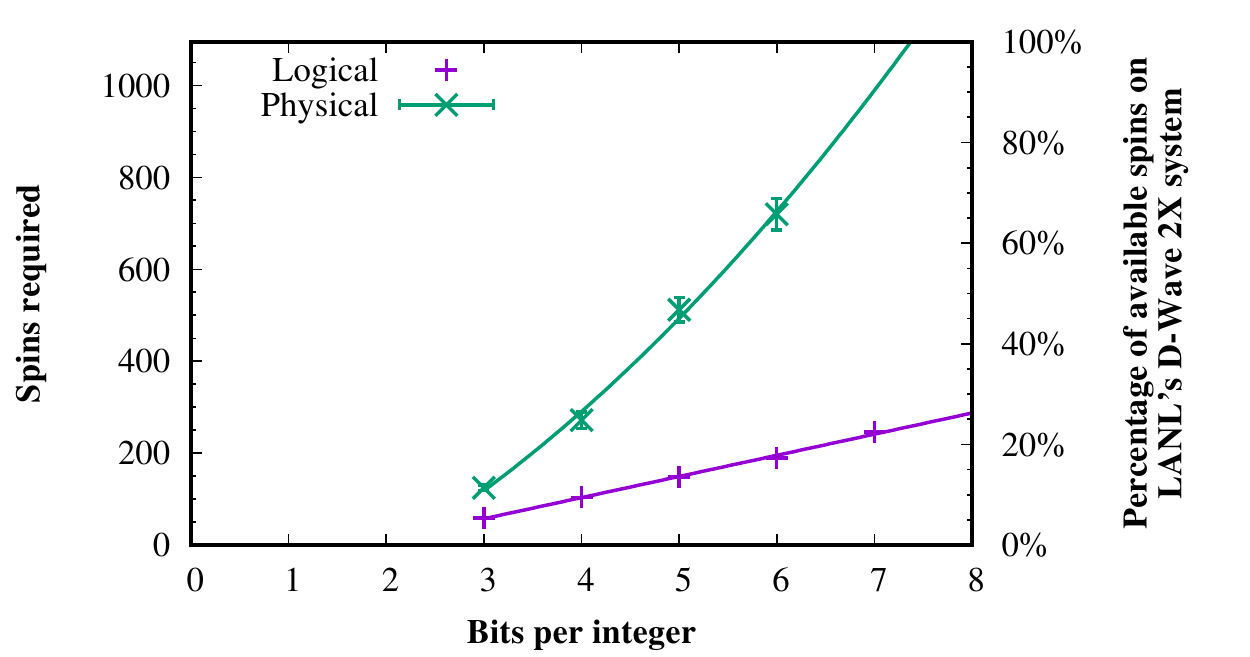}
  \caption{Cost in spins for ``\texttt{mult(P, Q, 6)}'' with different integer bit widths}
  \label{fig:factor-cost}
\end{figure}

\Cref{fig:factor-cost} presents the results of this study.
The number of logical spins for a given bit width does not change from
compilation to compilation, but the number of physical spins does
because it relies on a stochastic minor-embedding
algorithm~\cite{Cai2014:graph-minors}.  Consequently, the figure
includes error bars for physical spin counts.  Points represent
measurements; lines represent regressions.  The regression curves used
are $f_{\text{Log}}(x) = 46x - 80.6$ and $f_{\text{Phys}}(x) = 15.2x^2
+ 65.72x - 215.44$.  Both have a coefficient of determination $R^2 >
0.996$.  As \cref{fig:factor-cost} indicates, the number of physical
spins grows faster than the number of logical spins.  For integer
factorization, we run out of physical spins at 7~bits per integer.

%% file: conclusions.tex
\section{Conclusions}
\label{sec:conclusions}

Quantum annealers represent a radical departure from conventional
computer architectures.  Rather than perform a sequence of operations
that modify state (registers and memory), a quantum annealer performs
in hardware a particular type of NP-hard optimization problem.
Specifically, it finds a set of Boolean values (spins) that minimize a
real-valued, fixed-form function of a potentially large number of
variables.  In effect, a quantum annealer evaluates the function in
parallel for all $2^N$ possible inputs and reports in a fixed length
of time (microseconds) the best instance found.  The catch is that the
solution is not guaranteed to be optimal.

Although quantum annealers offer the potential for huge performance
gains through massive effective parallelism, programming them can be a
challenge.  At the lowest level, a program for a quantum annealer is
merely a list of coefficients for the aforementioned fixed-form
pseudo-Boolean function.  The question we sought to answer in this
work is, \textit{Can one express constraint logic programming in the
  form accepted by quantum-annealing hardware?}  The key insight we
make in answering that question is identifying an analogy between
expression minimization in an Ising-model Hamiltonian and unification
in constraint logic programming.

Based on that insight we implemented \qap, a compiler that, through a
sequence of transformations, converts Prolog programs into 2-local
Ising-model Hamiltonians, runs these on a \dw\ quantum annealer, and
reports the results in terms of program variables.  We draw the
following conclusions from our study:

\begin{enumerate}
  \item Despite the enormous semantic gap, it is indeed possible to
    automatically convert constraint logic programs, expressed in a
    subset of Prolog, to the solution to an optimization problem,
    expressed as coefficients to a 2-local Ising-model Hamiltonian.

  \item An important limiting factor is the number of spins needed to
    express even trivial constraint logic problems.  Our experimental
    platform, a \dwiix\ quantum annealer installed at Los Alamos
    National Laboratory, has 1095~spins.  One can think of those
    1095~spins as corresponding to roughly 136~bytes of memory, which
    need to hold all program inputs, outputs, and logic.

  \item End-to-end performance (i.e.,~including compilation time) is
    poor: a few seconds for even trivial CLP problems.  Although we
    expect these times to rise slowly with problem size and complexity
    we cannot confirm that hypothesis or compare it to classical
    implementations until we can execute programs sufficiently large
    so as to challenge classical CLP systems.
\end{enumerate}

Even considering the preceding shortcomings we remain optimistic about
the potential of exploiting the massive effective parallelism provided
by quantum annealers to accelerate the execution of constraint logic
programs.  Although the hardware is in its early generations, when
increased scale and other engineering improvements are put into place,
\qap\ will be ready to take advantage of these advances.

%% file: example.tex
\section{An end-to-end example}
\label{sec:example}

In this section we detail the complete \qap\ compilation and execution
process illustrated in \cref{fig:qap-flow}.  \Cref{fig:end-end-prolog}
presents a Prolog source file that defines a \texttt{bigger/2}
predicate, which we query with ``\texttt{bigger(Big, Little)}''.
Because the constraint logic works on free variables, this example
does not work with ordinary Prolog environments.  For instance,
\mbox{SWI-Prolog}~\cite{Wielemaker2012:swi-prolog} returns an
``\texttt{Arguments are not sufficiently instantiated}'' error.

\begin{figure}
  \centering
  \begin{minipage}[t]{6pc}
    \figrule
    \begin{lstlisting}[
        language=Prolog,
        columns=fullflexible,
        breaklines
      ]
bigger(X, Y) :-
    X > Y,
    X > 0,
    X < 10,
    Y > 6,
    Y < 15.
    \end{lstlisting}
    \figrule
  \end{minipage}
  \caption{Sample Prolog source file}
  \label{fig:end-end-prolog}
\end{figure}

\qap\ compiles \texttt{bigger/2} and the associated query into the
Verilog code shown in \cref{fig:end-end-verilog}.  Because the largest
integer literal appearing in the Prolog source code is~15, the Verilog
code uses 4-bit integers throughout.

\begin{figure}
  \centering
  \begin{minipage}[t]{17pc}
    \figrule
    \begin{lstlisting}[
        language=Verilog,
        columns=fullflexible,
        breaklines
      ]
module \bigger/2 (A, B, Valid);
  input [3:0] A;
  input [3:0] B;
  output Valid;
  wire [4:0] $v1;
  assign $v1[0] = A > B;
  assign $v1[1] = A > 4'd0;
  assign $v1[2] = A < 4'd10;
  assign $v1[3] = B > 4'd6;
  assign $v1[4] = B < 4'd15;
  assign Valid = &$v1;
endmodule

module Query (Big, Little, Valid);
  input [3:0] Big;
  input [3:0] Little;
  output Valid;
  wire $v1;
  \bigger/2 \bigger_xvLbZ/2 (Big, Little, $v1);
  assign Valid = &$v1;
endmodule
    \end{lstlisting}
    \figrule
  \end{minipage}
  \caption{Verilog code generated from \cref{fig:end-end-prolog} by \qap}
  \label{fig:end-end-verilog}
\end{figure}

\begin{figure}
  \centering
  \includegraphics[width=\linewidth]{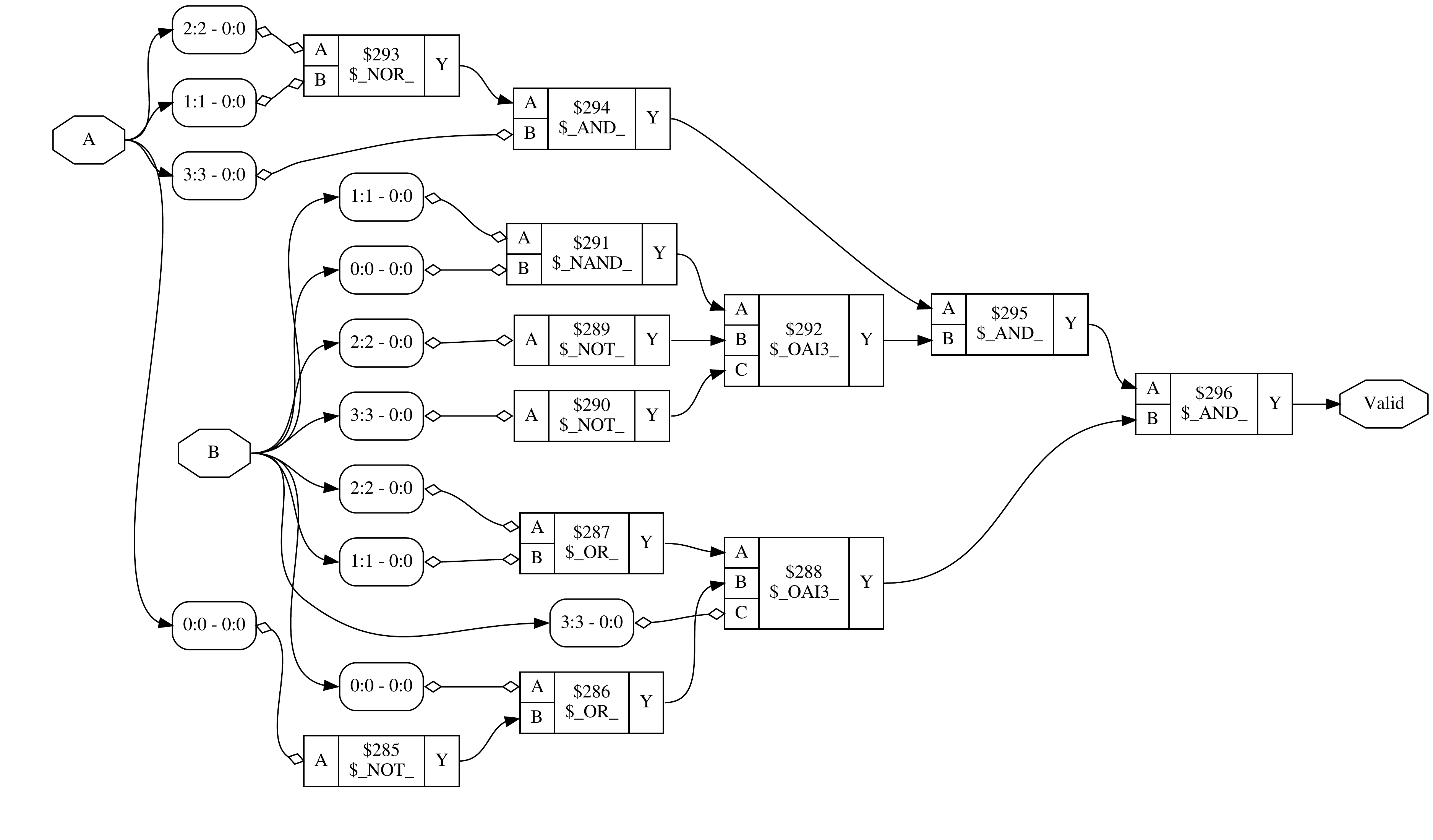}
  \caption{Visualization of the optimized netlist Yosys produced from \cref{fig:end-end-verilog}}
  \label{fig:end-end-edif}
\end{figure}

The EDIF netlist that Yosys produces from \cref{fig:end-end-verilog}
is a rather verbose \mbox{s-expression}.  Rather than present it here,
\cref{fig:end-end-edif} shows a visualization of the circuit that that
Yosys renders using Graphviz~\cite{Gansner2000:graphviz}.  The
notation that Yosys uses indicates how bits are renumbered as they
flow from one component to the next.  ``OAI3'' represents a 3-input
\bool{or}-\bool{and}-invert gate: $Y = \neg ((A \vee B) \wedge C)$.

edif2qmasm translates the circuit illustrated in
\cref{fig:end-end-edif} to the QMASM code listed in
\cref{fig:end-end-qmasm}.  In QMASM syntax, ``\texttt{=}'' specifies a
\defn{chain} (a strongly negative \Jij) and ``\texttt{<->}'' specifies
an \emph{alias} (multiple names for the same spin).  The translation
from EDIF constructs to QMASM constructs is nearly one-to-one, which
makes the process fairly straightforward.

\begin{figure}
  \figrule
  \begin{lstlisting}[
      basicstyle=\footnotesize,
      multicols=2,
      columns=fullflexible,
      alsoletter=!,
      morekeywords={!begin_macro,!end_macro,!use_macro,!include},
      morecomment={[l]\#},
      breaklines
    ]
!include <stdcell>

# bigger/2
!begin_macro id00007
  !use_macro NOT $id00008
  !use_macro OR $id00009
  !use_macro OR $id00010
  !use_macro OAI3 $id00011
  !use_macro NOT $id00012
  !use_macro NOT $id00013
  !use_macro NAND $id00014
  !use_macro OAI3 $id00015
  !use_macro NOR $id00016
  !use_macro AND $id00017
  !use_macro AND $id00018
  !use_macro AND $id00019
  $id00009.B = $id00008.Y
  $id00011.A = $id00010.Y
  $id00019.B = $id00011.Y
  $id00011.B = $id00009.Y
  Valid = $id00019.Y
  B[0] = $id00009.A 
  B[0] = $id00014.B 
  $id00009.A = $id00014.B  
  B[0] <-> $id00009.A
  B[1] = $id00014.A  
  B[1] = $id00010.B  
  $id00014.A = $id00010.B  
  B[1] <-> $id00014.A
  B[2] = $id00010.A  
  B[2] = $id00012.A  
  $id00010.A = $id00012.A  
  B[2] <-> $id00010.A
  B[3] = $id00013.A  
  B[3] = $id00011.C  
  $id00013.A = $id00011.C  
  B[3] <-> $id00013.A
  A[1] = $id00016.B  
  A[0] = $id00008.A  
  A[3] = $id00017.B  
  A[2] = $id00016.A  
  $id00017.A = $id00016.Y
  $id00015.C = $id00013.Y
  $id00018.B = $id00015.Y
  $id00019.A = $id00018.Y
  $id00018.A = $id00017.Y
  $id00015.B = $id00012.Y
  $id00015.A = $id00014.Y
!end_macro id00007

!begin_macro Query
  !use_macro id00007 $id00039
  $id00039.Valid = Valid  
  $id00039.B[1] = Little[1]  
  $id00039.B[0] = Little[0]  
  $id00039.B[3] = Little[3]  
  $id00039.B[2] = Little[2]  
  $id00039.A[2] = Big[2]  
  $id00039.A[3] = Big[3]  
  $id00039.A[0] = Big[0]  
  $id00039.A[1] = Big[1]  
!end_macro Query

!use_macro Query Query
  \end{lstlisting}
  \figrule
  \caption{QMASM code generated from the EDIF version of \cref{fig:end-end-edif} by edif2qmasm}
  \label{fig:end-end-qmasm}
\end{figure}

Finally, QMASM assigns each variable to one or more physical spins
that are available on the target hardware.  Because the process
contains some stochastic elements, \cref{fig:end-end-hamiltonian}
presents only one instance of a mapping to the hardware.  In the case
shown, the Hamiltonian uses 76~\hi\ and 101~\Jij.
\begin{figure}
  \figrule
  \fontsize{9}{13}\selectfont 
  $\mathcal{H}_{\text{Query}}(\bm{\sigma}) = \frac{5}{80} \sigma_{0} +
  \frac{1}{12} \sigma_{1} + \frac{1}{8} \sigma_{3} + \frac{1}{8}
  \sigma_{5} + \frac{5}{80} \sigma_{6} + \frac{5}{80} \sigma_{10} -
  \frac{1}{2} \sigma_{14} - \frac{1}{8} \sigma_{15} - \frac{1}{4}
  \sigma_{16} - \frac{1}{8} \sigma_{17} - \frac{1}{8} \sigma_{19} -
  \frac{1}{8} \sigma_{20} - \frac{1}{8} \sigma_{22} - \frac{1}{8}
  \sigma_{23} + \frac{5}{80} \sigma_{96} + \frac{1}{12} \sigma_{97} +
  \frac{1}{2} \sigma_{99} + \frac{5}{80} \sigma_{100} + \sigma_{101} +
  \frac{1}{12} \sigma_{102} - \frac{5}{80} \sigma_{104} + \frac{5}{80}
  \sigma_{106} - \frac{5}{80} \sigma_{107} + \frac{5}{80} \sigma_{108}
  - \frac{5}{80} \sigma_{110} - \frac{1}{8} \sigma_{113} - \frac{1}{8}
  \sigma_{115} + \frac{1}{2} \sigma_{117} - \frac{5}{80} \sigma_{118}
  - \frac{1}{8} \sigma_{119} + \frac{5}{80} \sigma_{192} - \frac{1}{8}
  \sigma_{194} - \frac{1}{12} \sigma_{195} - \frac{1}{12} \sigma_{196}
  - \frac{1}{8} \sigma_{200} - \frac{1}{2} \sigma_{203} - \frac{1}{8}
  \sigma_{204} + \frac{1}{12} \sigma_{205} - \frac{1}{2} \sigma_{208}
  - \frac{1}{8} \sigma_{209} + \frac{1}{12} \sigma_{210} +
  \frac{1}{12} \sigma_{213} + \frac{1}{4} \sigma_{214} - \frac{1}{8}
  \sigma_{290} - \frac{1}{12} \sigma_{291} + \frac{1}{2} \sigma_{294}
  - \frac{1}{4} \sigma_{296} - \frac{3}{8} \sigma_{297} - \frac{5}{80}
  \sigma_{299} - \frac{5}{80} \sigma_{300} - \frac{1}{4} \sigma_{301}
  - \frac{1}{8} \sigma_{302} - \frac{1}{4} \sigma_{303} + \frac{1}{4}
  \sigma_{305} + \frac{1}{4} \sigma_{309} - \frac{1}{8} \sigma_{310} +
  \frac{1}{4} \sigma_{1} \sigma_{5} - \sigma_{3} \sigma_{5} -
  \sigma_{0} \sigma_{6} - \sigma_{5} \sigma_{13} - \frac{1}{2}
  \sigma_{6} \sigma_{14} - \frac{1}{2} \sigma_{10} \sigma_{14} -
  \sigma_{9} \sigma_{15} + \frac{1}{8} \sigma_{16} \sigma_{20} +
  \frac{1}{4} \sigma_{17} \sigma_{20} - \sigma_{19} \sigma_{20} -
  \sigma_{14} \sigma_{22} + \frac{1}{4} \sigma_{16} \sigma_{22} +
  \frac{3}{8} \sigma_{17} \sigma_{22} + \frac{1}{4} \sigma_{19}
  \sigma_{22} - \sigma_{15} \sigma_{23} + \frac{1}{2} \sigma_{16}
  \sigma_{23} - \sigma_{17} \sigma_{23} + \frac{1}{4} \sigma_{19}
  \sigma_{23} - \sigma_{0} \sigma_{96} - \sigma_{1} \sigma_{97} +
  \frac{1}{2} \sigma_{3} \sigma_{99} + \frac{1}{4} \sigma_{96}
  \sigma_{100} - \sigma_{97} \sigma_{102} + \frac{1}{2} \sigma_{99}
  \sigma_{102} - \sigma_{8} \sigma_{104} + \frac{1}{2} \sigma_{9}
  \sigma_{105} - \sigma_{10} \sigma_{106} - \sigma_{100} \sigma_{108}
  - \frac{1}{3} \sigma_{104} \sigma_{108} - \sigma_{106} \sigma_{108}
  - \frac{1}{3} \sigma_{107} \sigma_{108} + \sigma_{101} \sigma_{109}
  - \sigma_{104} \sigma_{110} - \frac{1}{3} \sigma_{106} \sigma_{110}
  - \sigma_{107} \sigma_{110} - \frac{1}{8} \sigma_{16} \sigma_{112} -
  \sigma_{19} \sigma_{115} - \sigma_{108} \sigma_{116} - \sigma_{114}
  \sigma_{116} - \sigma_{109} \sigma_{117} - \frac{1}{2} \sigma_{113}
  \sigma_{117} - \frac{1}{2} \sigma_{115} \sigma_{117} - \sigma_{110}
  \sigma_{118} - \sigma_{114} \sigma_{118} + \frac{1}{8} \sigma_{112}
  \sigma_{119} + \frac{1}{4} \sigma_{113} \sigma_{119} - \sigma_{115}
  \sigma_{119} - \sigma_{96} \sigma_{192} - \sigma_{97} \sigma_{193} -
  \sigma_{99} \sigma_{195} + \frac{1}{4} \sigma_{194} \sigma_{196} -
  \sigma_{195} \sigma_{196} - \sigma_{192} \sigma_{197} - \sigma_{194}
  \sigma_{198} + \frac{1}{4} \sigma_{104} \sigma_{200} - \sigma_{105}
  \sigma_{201} + \frac{1}{2} \sigma_{107} \sigma_{203} - \sigma_{200}
  \sigma_{204} - \sigma_{202} \sigma_{204} + \frac{1}{2} \sigma_{203}
  \sigma_{204} - \sigma_{200} \sigma_{205} - \sigma_{202} \sigma_{205}
  - \sigma_{112} \sigma_{208} - \sigma_{113} \sigma_{209} +
  \frac{1}{2} \sigma_{211} \sigma_{212} - \sigma_{205} \sigma_{213} -
  \frac{1}{2} \sigma_{208} \sigma_{213} - \sigma_{210} \sigma_{213} -
  \frac{1}{2} \sigma_{208} \sigma_{214} + \frac{1}{4} \sigma_{210}
  \sigma_{214} - \sigma_{211} \sigma_{214} - \sigma_{210} \sigma_{215}
  - \sigma_{192} \sigma_{288} - \sigma_{194} \sigma_{290} -
  \sigma_{195} \sigma_{291} + \frac{1}{4} \sigma_{288} \sigma_{293} -
  \sigma_{289} \sigma_{293} - \frac{1}{2} \sigma_{290} \sigma_{294} -
  \frac{1}{2} \sigma_{291} \sigma_{294} + \frac{1}{4} \sigma_{288}
  \sigma_{295} - \sigma_{289} \sigma_{295} - \sigma_{201} \sigma_{297}
  - \sigma_{203} \sigma_{299} + \frac{1}{8} \sigma_{296} \sigma_{300}
  + \frac{3}{8} \sigma_{297} \sigma_{300} - \sigma_{299} \sigma_{300}
  + \frac{1}{8} \sigma_{293} \sigma_{301} - \sigma_{296} \sigma_{301}
  + \frac{1}{2} \sigma_{297} \sigma_{301} + \frac{1}{8} \sigma_{299}
  \sigma_{301} - \sigma_{294} \sigma_{302} + \frac{1}{4} \sigma_{296}
  \sigma_{302} - \frac{1}{8} \sigma_{295} \sigma_{303} + \frac{1}{8}
  \sigma_{296} \sigma_{303} + \frac{1}{2} \sigma_{297} \sigma_{303} +
  \frac{1}{4} \sigma_{299} \sigma_{303} - \sigma_{209} \sigma_{305} -
  \frac{1}{2} \sigma_{301} \sigma_{309} - \sigma_{305} \sigma_{309} -
  \sigma_{302} \sigma_{310} - \frac{1}{2} \sigma_{305} \sigma_{310}$
  \figrule
  \caption{Final Hamiltonian executed on LANL's \dwiix}
  \label{fig:end-end-hamiltonian}
\end{figure}
The \texttt{Big} variable in the Prolog query is mapped to the bit
string $\sigma_{198} \, \sigma_{13} \, \sigma_{193} \, \sigma_{212}$,
and the \texttt{Little} variable in the query is mapped to the bit
string $\sigma_{9} \, \sigma_{197} \, \sigma_{8} \, \sigma_{215}$
(both big-endian).  The \texttt{Valid} bit
(\cref{fig:end-end-verilog}) is mapped to $\sigma_{109}$.

We ran the \cref{fig:end-end-hamiltonian} Hamiltonian on LANL's
\dwiix\ system, specifying that we wanted 1000~samples of the
$\bm{\sigma}$ vector and an annealing time of 20\textmu{}s.  The
\dwiix\ found exactly the three valid solutions: ``\texttt{Big = 8,
  Little = 7}'' (139 instances), ``\texttt{Big = 9, Little = 7}'' (137
instances), and ``\texttt{Big = 9, Little = 8}'' (226 instances).
QMASM automatically rejected the remaining instances.  (It is not
uncommon for a quantum annealer to return an invalid solution, which
is analogous to a classical optimization algorithm getting stuck in a
local minimum.)  Although only 20\,ms were spent performing quantum
annealing, a total of 200\,ms of real time was spent on the \dwiix\@.
This includes various requisite pre- and post-processing tasks.

The total end-to-end time, including not only the time spent on the
\dwiix\ but also the time spent in \qap, Yosys, edif2qmasm, QMASM, the
local filesystem, and various networks between the development
workstation and the \dwiix, averages $3.2 \pm 1.1$\,s over 100 trials.
Clearly, a problem like the one used in this exercise is
insufficiently complex to overcome the numerous overheads and observe
an increase in performance over what can readily be achieved
classically.

%% file: tplp2017.bbl
\begin{thebibliography}{}

\bibitem[\protect\citeauthoryear{Barahona}{Barahona}{1982}]{Baharona1982:ising-np}
{\sc Barahona, F.} 1982.
\newblock On the computational complexity of {I}sing spin glass models.
\newblock {\em Journal of Physics~A: Mathematical and General\/}~{\em
  15,\/}~10, 3241.

\bibitem[\protect\citeauthoryear{{Berkeley Logic Synthesis and Verification
  Group}}{{Berkeley Logic Synthesis and Verification
  Group}}{2016}]{Berkeley2006:abc}
{\sc {Berkeley Logic Synthesis and Verification Group}}. 2016.
\newblock {ABC}: A system for sequential synthesis and verification.
\newblock Available from \url{http://www.eecs.berkeley.edu/~alanmi/abc/}.

\bibitem[\protect\citeauthoryear{Bravyi, Bessen, and Terhal}{Bravyi
  et~al\mbox{.}}{2006}]{Bravyi2006:stoqma}
{\sc Bravyi, S.}, {\sc Bessen, A.~J.}, {\sc and} {\sc Terhal, B.~M.} 2006.
\newblock {M}erlin-{A}rthur games and stoquastic complexity.
\newblock {\em arXiv:quant-ph/0611021v2\/}.

\bibitem[\protect\citeauthoryear{Bravyi and Hastings}{Bravyi and
  Hastings}{2017}]{Bravyi2017:stoqma}
{\sc Bravyi, S.} {\sc and} {\sc Hastings, M.} 2017.
\newblock On complexity of the quantum {I}sing model.
\newblock {\em Communications in Mathematical Physics\/}~{\em 349,\/}~1 (Jan.),
  1--45.

\bibitem[\protect\citeauthoryear{Brayton and Mishchenko}{Brayton and
  Mishchenko}{2010}]{Brayton2010:abc}
{\sc Brayton, R.} {\sc and} {\sc Mishchenko, A.} 2010.
\newblock {ABC}: An academic industrial-strength verification tool.
\newblock In {\em Proceedings of the 22nd International Conference on Computer
  Aided Verification}, {T.~Touili}, {B.~Cook}, {and} {P.~Jackson}, Eds.
  Springer Berlin Heidelberg, Edinburgh, United Kingdom, 24--40.

\bibitem[\protect\citeauthoryear{Bunyk, Hoskinson, Johnson, Tolkacheva,
  Altomare, Berkley, Harris, Hilton, Lanting, Przybysz, and Whittaker}{Bunyk
  et~al\mbox{.}}{2014}]{Bunyk2014:dwave-arch}
{\sc Bunyk, P.~I.}, {\sc Hoskinson, E.~M.}, {\sc Johnson, M.~W.}, {\sc
  Tolkacheva, E.}, {\sc Altomare, F.}, {\sc Berkley, A.~J.}, {\sc Harris, R.},
  {\sc Hilton, J.~P.}, {\sc Lanting, T.}, {\sc Przybysz, A.~J.}, {\sc and} {\sc
  Whittaker, J.} 2014.
\newblock Architectural considerations in the design of a superconducting
  quantum annealing processor.
\newblock {\em {IEEE} {T}ransactions on {A}pplied {S}uperconductivity\/}~{\em
  24,\/}~4 (Aug.), 1--10.

\bibitem[\protect\citeauthoryear{Cai, Macready, and Roy}{Cai
  et~al\mbox{.}}{2014}]{Cai2014:graph-minors}
{\sc Cai, J.}, {\sc Macready, B.}, {\sc and} {\sc Roy, A.} 2014.
\newblock A practical heuristic for finding graph minors.
\newblock {\em arXiv:1406.2741 [quant-ph]\/}.

\bibitem[\protect\citeauthoryear{Choi}{Choi}{2008}]{Choi2008:minor-embed}
{\sc Choi, V.} 2008.
\newblock Minor-embedding in adiabatic quantum computation: {I}.~the parameter
  setting problem.
\newblock {\em Quantum Information Processing\/}~{\em 7,\/}~5 (Oct.), 193--209.

\bibitem[\protect\citeauthoryear{Cormen, Leiserson, Rivest, and Stein}{Cormen
  et~al\mbox{.}}{2001}]{Cormen2001:algorithms}
{\sc Cormen, T.~H.}, {\sc Leiserson, C.~E.}, {\sc Rivest, R.~L.}, {\sc and}
  {\sc Stein, C.} 2001.
\newblock {\em Introduction to Algorithms\/}, 2nd ed.
\newblock MIT Press.

\bibitem[\protect\citeauthoryear{Cross, Bishop, Smolin, and Gambetta}{Cross
  et~al\mbox{.}}{2017}]{Cross2017:openqasm}
{\sc Cross, A.~W.}, {\sc Bishop, L.~S.}, {\sc Smolin, J.~A.}, {\sc and} {\sc
  Gambetta, J.~M.} 2017.
\newblock Open quantum assembly language.
\newblock {\em arXiv:1707.03429 [quant-ph]\/}.

\bibitem[\protect\citeauthoryear{D-Wave Systems, Inc.}{D-Wave Systems,
  Inc.}{2017a}]{D-Wave2016:sapi-python}
D-Wave Systems, Inc. 2017a.
\newblock {\em Developer Guide for {P}ython}.
\newblock D-Wave Systems, Inc., Burnaby, British Columbia, Canada.

\bibitem[\protect\citeauthoryear{D-Wave Systems, Inc.}{D-Wave Systems,
  Inc.}{2017b}]{D-Wave:toq}
D-Wave Systems, Inc. 2017b.
\newblock {\em {T}o{Q} Overview}.
\newblock D-Wave Systems, Inc., Burnaby, British Columbia, Canada.
\newblock ToQ documentation, qOp version~2.3.1.

\bibitem[\protect\citeauthoryear{Dahl}{Dahl}{2014}]{Dahl2014:deqo}
{\sc Dahl, E.~D.} 2014.
\newblock {\em {D}eqo: A Direct Embedding Quantum Optimizer}.
\newblock D-Wave Systems, Inc.

\bibitem[\protect\citeauthoryear{Dutra}{Dutra}{2010}]{Dutra2010:clp-tut}
{\sc Dutra, I.} 2010.
\newblock Constraint logic programming: A short tutorial.
\newblock Available from \url{https://www.dcc.fc.up.pt/~ines/talks/clp-v1.pdf}.

\bibitem[\protect\citeauthoryear{Farhi and Gutmann}{Farhi and
  Gutmann}{1998}]{Farhi1998:adiabatic}
{\sc Farhi, E.} {\sc and} {\sc Gutmann, S.} 1998.
\newblock Analog analogue of a digital quantum computation.
\newblock {\em Physical Review~A\/}~{\em 57}, 2403--2406.

\bibitem[\protect\citeauthoryear{Feynman}{Feynman}{1986}]{Feynman1986:quantum-gates}
{\sc Feynman, R.~P.} 1986.
\newblock Quantum mechanical computers.
\newblock {\em {F}oundations of {P}hysics\/}~{\em 16,\/}~6 (June), 507--531.

\bibitem[\protect\citeauthoryear{Finnila, Gomez, Sebenik, Stenson, and
  Doll}{Finnila et~al\mbox{.}}{1994}]{Finnila1994:quant-anneal}
{\sc Finnila, A.~B.}, {\sc Gomez, M.~A.}, {\sc Sebenik, C.}, {\sc Stenson, C.},
  {\sc and} {\sc Doll, J.~D.} 1994.
\newblock Quantum annealing: A new method for minimizing multidimensional
  functions.
\newblock {\em Chemical Physics Letters\/}~{\em 219,\/}~5, 343--348.

\bibitem[\protect\citeauthoryear{{Fujitsu Ltd.}}{{Fujitsu
  Ltd.}}{2017}]{Fujitsu2017:dig-anneal}
{\sc {Fujitsu Ltd.}} 2017.
\newblock Quantum computing and {AI} start a new era.
\newblock {\em Fujitsu Journal\/}.

\bibitem[\protect\citeauthoryear{Gansner and North}{Gansner and
  North}{2000}]{Gansner2000:graphviz}
{\sc Gansner, E.~R.} {\sc and} {\sc North, S.~C.} 2000.
\newblock An open graph visualization system and its applications to software
  engineering.
\newblock {\em Software---Practice and Experience\/}~{\em 30,\/}~11 (Sept.),
  1203--1233.

\bibitem[\protect\citeauthoryear{Green, Lumsdaine, Ross, Selinger, and
  Valiron}{Green et~al\mbox{.}}{2013}]{Green2013:quipper}
{\sc Green, A.~S.}, {\sc Lumsdaine, P.~L.}, {\sc Ross, N.~J.}, {\sc Selinger,
  P.}, {\sc and} {\sc Valiron, B.} 2013.
\newblock Quipper: A scalable quantum programming language.
\newblock In {\em Proceedings of the 34th ACM SIGPLAN Conference on Programming
  Language Design and Implementation}. ACM, New York, New York, USA, 333--342.

\bibitem[\protect\citeauthoryear{Grover}{Grover}{1996}]{Grover1996:search}
{\sc Grover, L.~K.} 1996.
\newblock A fast quantum mechanical algorithm for database search.
\newblock In {\em Proceedings of the Twenty-eighth Annual ACM Symposium on
  Theory of Computing}. ACM, New York, NY, USA, 212--219.

\bibitem[\protect\citeauthoryear{Heim, Brown, Wecker, and Troyer}{Heim
  et~al\mbox{.}}{2017}]{Heim2017:adiabatic-tsp}
{\sc Heim, B.}, {\sc Brown, E.~W.}, {\sc Wecker, D.}, {\sc and} {\sc Troyer,
  M.} 2017.
\newblock Designing adiabatic quantum optimization: A case study for the
  traveling salesman problem.
\newblock {\em arXiv:1702.06248v1 [quant-ph]\/}.

\bibitem[\protect\citeauthoryear{Hen and Young}{Hen and
  Young}{2011}]{Hen2011:exp-complex}
{\sc Hen, I.} {\sc and} {\sc Young, A.~P.} 2011.
\newblock Exponential complexity of the quantum adiabatic algorithm for certain
  satisfiability problems.
\newblock {\em Physical Review~E\/}~{\em 84}, 061152.

\bibitem[\protect\citeauthoryear{James, Ortiz, and Sabry}{James
  et~al\mbox{.}}{2011}]{James2011:qc-prolog}
{\sc James, R.~P.}, {\sc Ortiz, G.}, {\sc and} {\sc Sabry, A.} 2011.
\newblock Quantum computing over finite fields.
\newblock {\em arXiv:1101.3764 [quant-ph]\/}.

\bibitem[\protect\citeauthoryear{JavadiAbhari, Patil, Kudrow, Heckey, Lvov,
  Chong, and Martonosi}{JavadiAbhari
  et~al\mbox{.}}{2015}]{JavadiAbhari2015:scaffcc}
{\sc JavadiAbhari, A.}, {\sc Patil, S.}, {\sc Kudrow, D.}, {\sc Heckey, J.},
  {\sc Lvov, A.}, {\sc Chong, F.~T.}, {\sc and} {\sc Martonosi, M.} 2015.
\newblock {S}caff{CC}: Scalable compilation and analysis of quantum programs.
\newblock {\em Parallel Computing\/}~{\em 45}, 2--17.

\bibitem[\protect\citeauthoryear{Johnson, Amin, Gildert, Lanting, Hamze,
  Dickson, Harris, Berkley, Johansson, Bunyk, Chapple, Enderud, Hilton, Karimi,
  Ladizinsky, Ladizinsky, Oh, Perminov, Rich, Thom, Tolkacheva, Truncik,
  Uchaikin, Wang, Wilson, and Rose}{Johnson
  et~al\mbox{.}}{2011}]{Johnson2011:quant-ann}
{\sc Johnson, M.~W.}, {\sc Amin, M. H.~S.}, {\sc Gildert, S.}, {\sc Lanting,
  T.}, {\sc Hamze, F.}, {\sc Dickson, N.}, {\sc Harris, R.}, {\sc Berkley,
  A.~J.}, {\sc Johansson, J.}, {\sc Bunyk, P.}, {\sc Chapple, E.~M.}, {\sc
  Enderud, C.}, {\sc Hilton, J.~P.}, {\sc Karimi, K.}, {\sc Ladizinsky, E.},
  {\sc Ladizinsky, N.}, {\sc Oh, T.}, {\sc Perminov, I.}, {\sc Rich, C.}, {\sc
  Thom, M.~C.}, {\sc Tolkacheva, E.}, {\sc Truncik, C. J.~S.}, {\sc Uchaikin,
  S.}, {\sc Wang, J.}, {\sc Wilson, B.}, {\sc and} {\sc Rose, G.} 2011.
\newblock Quantum annealing with manufactured spins.
\newblock {\em Nature\/}~{\em 473,\/}~7346 (May), 194--198.

\bibitem[\protect\citeauthoryear{Kadowaki and Nishimori}{Kadowaki and
  Nishimori}{1998}]{Kadowaki1998:quant-anneal}
{\sc Kadowaki, T.} {\sc and} {\sc Nishimori, H.} 1998.
\newblock Quantum annealing in the transverse {I}sing model.
\newblock {\em {P}hysical {R}eview~{E}\/}~{\em 58}, 5355--5363.

\bibitem[\protect\citeauthoryear{Kahn, {La Fontaine}, and Lau}{Kahn
  et~al\mbox{.}}{2000}]{Kahn2000:edif}
{\sc Kahn, H.}, {\sc {La Fontaine}, R.}, {\sc and} {\sc Lau, R.} 2000.
\newblock Electronic design interchange format ({EDIF}): Part~2: Version~4~0~0.
\newblock Standard IEC 61690-2:2000, International Electrotechnical Commission,
  Manchester, United Kingdom. Jan.

\bibitem[\protect\citeauthoryear{Kaminsky, Lloyd, and Orlando}{Kaminsky
  et~al\mbox{.}}{2004}]{Kaminsky2004:adiabatic}
{\sc Kaminsky, W.~M.}, {\sc Lloyd, S.}, {\sc and} {\sc Orlando, T.~P.} 2004.
\newblock Scalable superconducting architecture for adiabatic quantum
  computation.
\newblock {\em arXiv:quant-ph/0403090v2\/}.

\bibitem[\protect\citeauthoryear{Kirkpatrick, Gelatt, and Vecchi}{Kirkpatrick
  et~al\mbox{.}}{1983}]{Kirkpatrick1983:sim-anneal}
{\sc Kirkpatrick, S.}, {\sc Gelatt, C.~D.}, {\sc and} {\sc Vecchi, M.~P.} 1983.
\newblock Optimization by simulated annealing.
\newblock {\em Science\/}~{\em 220,\/}~4598 (May), 671--680.

\bibitem[\protect\citeauthoryear{Knight}{Knight}{2017}]{Knight2017:ibm-50q}
{\sc Knight, W.} 2017.
\newblock {IBM} raises the bar with a 50-qubit quantum computer.
\newblock {\em MIT Technology Review\/}.

\bibitem[\protect\citeauthoryear{Lucas}{Lucas}{2014}]{Lucas2014:np-probs}
{\sc Lucas, A.} 2014.
\newblock {I}sing formulations of many {NP} problems.
\newblock {\em Frontiers in Physics\/}~{\em 2}, 5.

\bibitem[\protect\citeauthoryear{{Microsoft Corp.}}{{Microsoft
  Corp.}}{2017}]{Microsoft2017:q-sharp}
{\sc {Microsoft Corp.}} 2017.
\newblock The {Q\#} progamming language.
\newblock Available from
  \url{https://docs.microsoft.com/en-us/quantum/quantum-qr-intro}.

\bibitem[\protect\citeauthoryear{Nethercote, Stuckey, Becket, Brand, Duck, and
  Tack}{Nethercote et~al\mbox{.}}{2007}]{Nethercote2007:minizinc}
{\sc Nethercote, N.}, {\sc Stuckey, P.~J.}, {\sc Becket, R.}, {\sc Brand, S.},
  {\sc Duck, G.~J.}, {\sc and} {\sc Tack, G.} 2007.
\newblock {M}ini{Z}inc: Towards a standard {CP} modelling language.
\newblock In {\em Proceedings of the 13th International Conference on
  Principles and Practice of Constraint Programming}, {C.~Bessi{\`e}re}, Ed.
  Springer, Berlin, Heidelberg, 529--543.

\bibitem[\protect\citeauthoryear{Pakin}{Pakin}{2016}]{Pakin2016:qmasm}
{\sc Pakin, S.} 2016.
\newblock A quantum macro assembler.
\newblock In {\em Proceedings of the 2016 IEEE High Performance Extreme
  Computing Conference (HPEC)}.

\bibitem[\protect\citeauthoryear{Robinson}{Robinson}{1965}]{Robinson1965:unification}
{\sc Robinson, J.~A.} 1965.
\newblock A machine-oriented logic based on the resolution principle.
\newblock {\em Journal of the ACM\/}~{\em 12,\/}~1 (Jan.), 23--41.

\bibitem[\protect\citeauthoryear{Shor}{Shor}{1999}]{Shor1999:factoring}
{\sc Shor, P.~W.} 1999.
\newblock Polynomial-time algorithms for prime factorization and discrete
  logarithms on a quantum computer.
\newblock {\em SIAM Review\/}~{\em 41,\/}~2, 303--332.

\bibitem[\protect\citeauthoryear{Smith, Curtis, and Zeng}{Smith
  et~al\mbox{.}}{2017}]{Smith2017:quil}
{\sc Smith, R.~S.}, {\sc Curtis, M.~J.}, {\sc and} {\sc Zeng, W.~J.} 2017.
\newblock A practical quantum instruction set architecture.
\newblock {\em arXiv:1608.03355 [quant-ph]\/}.

\bibitem[\protect\citeauthoryear{Thomas and Moorby}{Thomas and
  Moorby}{2002}]{Thomas2002:verilog}
{\sc Thomas, D.} {\sc and} {\sc Moorby, P.} 2002.
\newblock {\em The {V}erilog Hardware Description Language\/}, 5th ed.
\newblock Springer.

\bibitem[\protect\citeauthoryear{Triska}{Triska}{2012}]{Triska2012:clpfd}
{\sc Triska, M.} 2012.
\newblock The finite domain constraint solver of {SWI}-{P}rolog.
\newblock In {\em Proceedings of the Eleventh International Symposium on
  Functional and Logic Programming (FLOPS~2012)}. Number 7294 in Lecture Notes
  in Computer Science. Kobe, Japan, 307--316.

\bibitem[\protect\citeauthoryear{Wecker and Svore}{Wecker and
  Svore}{2014}]{Wecker2014:liquid}
{\sc Wecker, D.} {\sc and} {\sc Svore, K.~M.} 2014.
\newblock {LIQ$Ui|\rangle$}: A software design architecture and domain-specific
  language for quantum computing.
\newblock {\em arXiv:1402.4467 [quant-ph]\/}.

\bibitem[\protect\citeauthoryear{Wielemaker, Schrijvers, Triska, and
  Lager}{Wielemaker et~al\mbox{.}}{2012}]{Wielemaker2012:swi-prolog}
{\sc Wielemaker, J.}, {\sc Schrijvers, T.}, {\sc Triska, M.}, {\sc and} {\sc
  Lager, T.} 2012.
\newblock {SWI}-{P}rolog.
\newblock {\em Theory and Practice of Logic Programming\/}~{\em 12,\/}~1--2
  (Jan.), 67--96.

\bibitem[\protect\citeauthoryear{Wolf and Glaser}{Wolf and
  Glaser}{2013}]{Wolf2013:yosys}
{\sc Wolf, C.} {\sc and} {\sc Glaser, J.} 2013.
\newblock {Y}osys---a free {V}erilog synthesis suite.
\newblock In {\em Proceedings of the 21st Austrian Workshop on Microelectronics
  (Austrochip~2013)}. Linz, Austria.

\bibitem[\protect\citeauthoryear{Yamaoka, Yoshimura, Hayashi, Okuyama, Aoki,
  and Mizuno}{Yamaoka et~al\mbox{.}}{2016}]{Yamaoka2016:cmos-anneal}
{\sc Yamaoka, M.}, {\sc Yoshimura, C.}, {\sc Hayashi, M.}, {\sc Okuyama, T.},
  {\sc Aoki, H.}, {\sc and} {\sc Mizuno, H.} 2016.
\newblock A 20k-spin {I}sing chip to solve combinatorial optimization problems
  with {CMOS} annealing.
\newblock {\em IEEE Journal of Solid-State Circuits\/}~{\em 51,\/}~1 (Jan.),
  303--309.

\end{thebibliography}
